\documentclass[aip,pof,amsmath,amssymb,preprint]{revtex4-1}
 \usepackage{graphicx}
 \usepackage{hyperref,rotating,bm,amsmath,tabularx,longtable}
\usepackage{amssymb}
\setcitestyle{authoryear}
\begin{document}

%\begin{frontmatter}



\title{Flow patterns and heat transfer around six in-line circular cylinders at low Reynolds number}

\author{Francesco Fornarelli}
\email{f.fornarelli@gmail.com}
\author{Paolo Oresta}
\author{Antonio Lippolis}
%\address[label1]{Dipartimento di Ingegneria dell'Innovazione, Universit\`a del Salento, 73100 Lecce, Italy.}
\affiliation{Dipartimento di Meccanica, Matematica e Management, Politecnico di Bari, 70126 Bari, Italy.}
%\address[label1]{Department of Engineering for Innovation, University of Salento, 73100 Lecce, Italy.}
%\address[label2]{Department of Mathematics, Mechanics and Management, Polytechnic of Bari, 70126 Bari, Italy.}
% Do not enter received and revised dates. These will be entered by the editorial office.
%\date{\today}
%\setcounter{page}{1}

\begin{abstract}

The flow field and the heat transfer around six in-line iso-thermal circular
cylinders has been studied by mean of numerical simulations.  Two values of
the center to center spacing ($s=3.6d$ and $4d$, where $d$ is the cylinder
diameter) at Reynolds number of $100$ and Prandtl number of $0.7$ has been
investigated. Similarly to the in-line two cylinder configuration, in this
range a transition in the flow and in the heat transfer occurs. Two different
flow patterns have been identified: the stable shear layer (SSL) mode and the
shear layer secondary vortices (SLSV) mode, at $3.6$ and $4$ spacing ratio
($s/d$), respectively. At $s/d=3.6$ the flow pattern causes the entrainment of
cold fluid on the downstream cylinders enhancing the heat transfer. On the
other hand at $s/d=4$ two stable opposite shear layer prevent the cold fluid
entrainment over the downstream cylinders reducing their heat exchange.  The
overall time average heat transfer of the array is enhanced up to 25\%
decreasing the spacing ratio from $4$ to $3.6$.  Furthermore, it is found that
the increased heat transfer is related to the phase shift between the Nusselt
time series of successive cylinders.  

\end{abstract}

i%\begin{keyword}
%Flow patterns \sep
%forced convection \sep
%circular cylinders \sep
%low Reynolds number \sep
%DNS \sep
%heat transfer

%\end{keyword}
\maketitle
%\end{frontmatter}

% \linenumbers

\section{Introduction}
\label{sec:Intro}



%% References % % Following citation commands can be used in the body text: %
%Usage of \cite is as follows: %   \cite{key}          ==>>  [#] % \cite[chap.
%2]{key} ==>>  [#, chap. 2] %   \citet{key}         ==>>  Author [#]

The fluid flow around bluff bodies is often subject of interest in theoretical
and applied physics. 
Heat transfer around an array of circular cylinders involves a variety of
engineering applications such as rod structure of the nuclear reactors,
compact heat exchangers for electronic components and pin-fins heat exchangers
for micro-devices (\cite{Kosar2005}).  In the micro-devices the Reynolds number
ranges between $10 - 200$ and the flow is mainly two-dimensional
(\cite{FLM:7244044}). The flow characteristics play a key role in
understanding how the array is able to enhance or decrease the heat exchange,
the mixing layer or the unsteady body forces
(\cite{Chatterjee20091114,Lo20126916,Lu20123909,Fornarelli2009}).
\cite{Sreedharan2008} experiments, on different pin geometries, highlight that
the heat transfer enhancement is related to the pin wake characteristics. In
particular the vorticity dynamics and the vortex street formation around
multiple bodies is strictly related to the wake interaction
(\cite{Hanson2011367,Rinaldi2012167,Ziada1992271}).

\begin{figure} 
\centering
\includegraphics[width=0.65\textwidth]{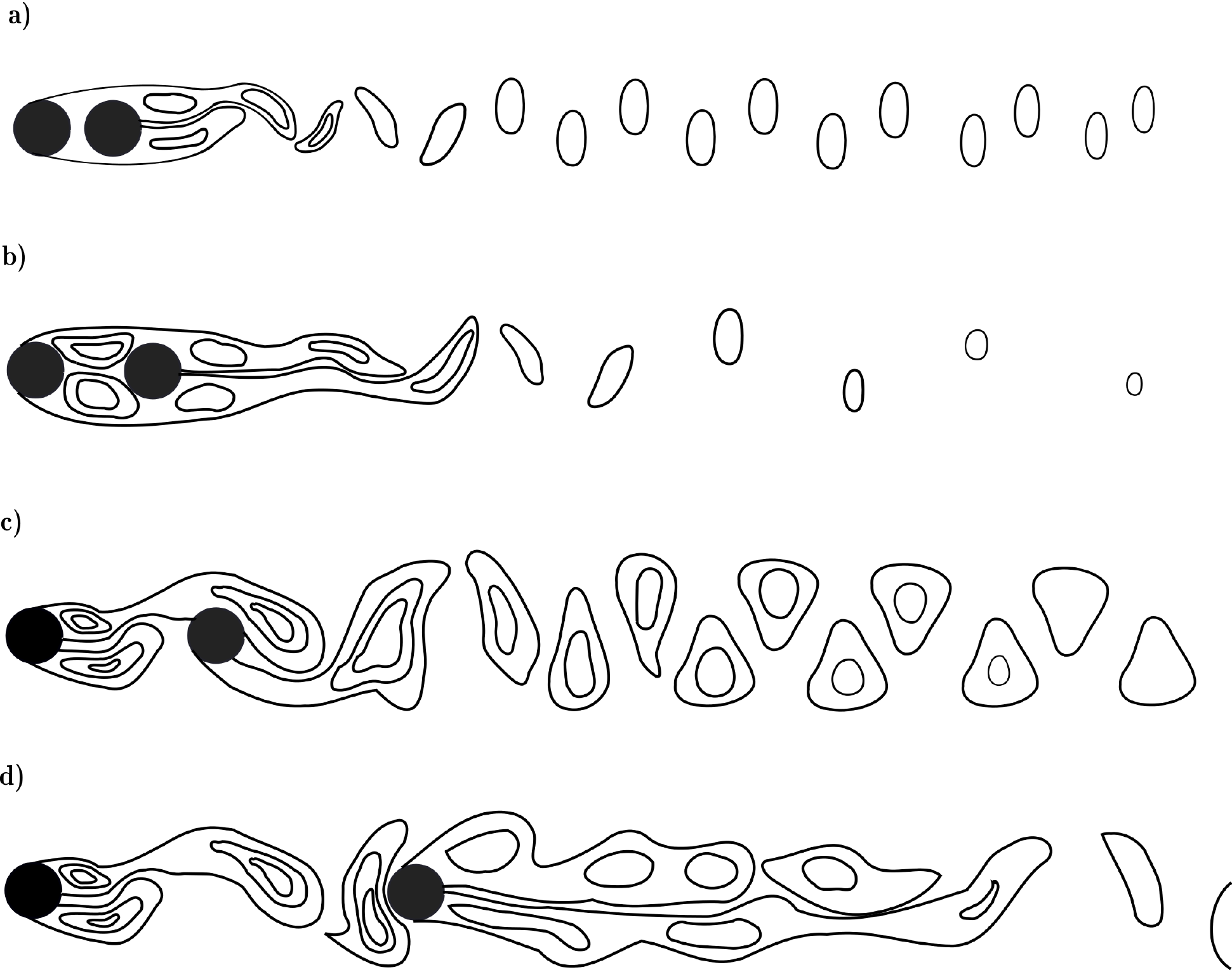} 
\caption{Schematic of the vorticity patterns for two tandem
cylinders varying the spacing ratio. a) SBB, b) SLR, c) SVS and d) SVF.}
\label{fig:modes} 
\end{figure}

The simple two cylinders configuration has been discussed extensively and it
represents a starting point to study the flow around multiple bluff bodies
(\cite{Meftah2013, FLM:8322063, Jester2003561, MITTAL2001717,
Prasanth2009731, FLD:FLD812, FLM:15867}). According to the literature, a
well-known {\it drag inversion separation} has been identified. Considering two
cylinders aligned with the flow direction and defining the distance between
their centers, the {\it drag inversion separation} represents the distance for
which the mean drag coefficient of the downstream cylinder switches from
positive to negative value due to the suction effect of the upstream cylinder.
A spacing larger than the {\it drag inversion separation} makes positive the
downstream cylinder drag coefficient and a vortex shedding in the gap between
the cylinders appears.  This flow is induced by the influence of both the
cylinders and it is called {\it wake interference} (\cite{Jester2003561}).
\cite{FLD:FLD812} studied numerically the flow around two tandem circular
cylinders at $Re = 100$ for the spacing ratio between $2$ and $10$ finding the
drag transition between $3.75$ and $4$.  In this configuration~\cite{FLM:15867}
identified three different flow patterns increasing the distance between the
cylinders called: single bluff body (SBB) mode, shear layer reattachment (SLR)
mode and synchronization of vortex shedding (SVS) mode. In SBB mode ($s/d < 1.2
- 1.8$ depending on Re) the two cylinders influence the flow field like a
single bluff body. In SLR mode ($1.2 - 1.8 < s/d < 3.4 - 3.8$) the vortices
shed by the upstream cylinder reattach at the downstream cylinder surface and
evolve in a von Karman vortex street. For larger $s/d$ than a critical value
both the cylinders shed vortices at the same frequency and this mode is called
SVS.  Even though~\cite{FLM:15867} investigated the flow at $850<Re<1900$,
recently \cite{PhysRevE.81.036305} founds, at low Reynolds numbers ($Re=60, 80,
100$), the same flow patterns; moreover, at larger spacing ratio than the
above-mentioned critical one, a new mode called the secondary vortex formation
(SVF) has been discovered. The SVF mode is characterized by secondary vortex
street in the wake of the downstream cylinder. A schematic of the vorticity
patterns of the above-mentioned modes is reported in fig.\ref{fig:modes} as
described in~\cite{PhysRevE.81.036305}.

  The flow dynamics of two cylinders highlight the strong dependence of
instability by the geometry and the flow conditions, therefore its relation to
multiple bodies configuration is not straight forward.  The two cylinder
configuration help the scientists and engineers to better understand the
physical mechanisms that influence the bluff bodies -- flow interaction, but
practical engineering applications, i.e.  tube bundle in heat exchanger,
involve numerous bluff bodies arranged in different array. In literature
several works extend the two cylinder configuration to multiple array, in
particular the six in-line configuration has been studied.  More insight is
provided by~\cite{Liang2009950},~\cite{FLM:8687985} and~\cite{Bao2012118} that
studied the flow characteristics of six in-line circular or square cylinders
at low Reynolds number. They focused on the near wake flow structures omitting
the analysis of the far wake structures.  \cite{Liang2009950} founds the
critical spacing between $3.6$  and $4$ of the center to center distance for
which the drag and lift coefficients of each cylinder change significantly.  

In this paper we made richer the six in-line circular cylinders results,
investigating by means of frequency analysis the time dependent cylinder
forces and flow velocities showing the pattern topology of the far wake.  Two
modes called stable shear layer (SSL) and shear layer secondary vortices
(SLSV) are found at $3.6$ and $4$ spacing ratio, respectively.  Farther the
cylinders has been heated and maintained at a constant temperature in order to
study the effects of the fluid flow on the heat transfer.  The forced
convection is considered. In this case the flow is not influenced by the
temperature since the buoyancy force is much smaller than the convective
force.  Considering the heat and mass transfer without buoyancy, two
dimensionless control parameters exist, the Reynolds ($Re$) and the Prandtl
($Pr$) number. They are defined as: 

\begin{align} \label{eq:RePr}
Re=\frac{U_0 d}{\nu} \quad & Pr=\frac{\nu}{\kappa}
%Ri=\frac{Gr}{Re^2} 
\end{align} 

where $U_0$, $d$, $\nu$ and $\kappa$ are, respectively, the inflow velocity,
the cylinder diameter, the kinematic viscosity of the fluid and its thermal
diffusivity.  In these hypotheses  the temperature is a scalar quantity
transported by the flow field with a proper diffusivity term that depends on
the Prandtl number. In this paper we used air as working fluid with a Prandtl
number at $20\,^{\circ}{\rm C}$ of $0.7$.  In~\cite{Harimi2012309} the
numerical simulation of forced convection at $Re=100-200$ with two in-line
isothermal cylinders shows an heat transfer enhancement increasing the spacing
ratio. Increasing the spacing ratio the heat exchange is enhanced especially
at the transition, $s/d=3.6-4$. It is worth to note that also the local
Nusselt number distribution on the surface of the downstream cylinder changes
in the transition range (\cite{Mahir20081309}) due to the shedding vortices in
the gap between the cylinders.  The results of the six in-line cylinders
configuration reveal a clear evidence that, on the contrary of the two tandem
cylinder configuration, the overall heat transfer increases up to the $25\%$
diminishing the spacing ratio from $4$ to $3.6$.

This paper is organized as follows.  First the numerical method is presented
and validated reproducing the flow around a single cylinder and a two tandem
circular cylinders configuration according to the literature.  The results of
the six circular cylinders in-line configuration around the critical spacing
is presented investigating the near and the far wake flow characteristics.
Finally the heat transfer around six in-line cylinders in the case of forced
convection at Prandtl number of $0.7$ is reported. 

\section {Governing equations and numerical method} 

The dimensionless two-dimensional incompressible Newtonian Navier-Stokes
reads:

\begin{equation}\label{eq:NS1} 
\frac{\partial{\mathbf u}}{\partial t}+{\mathbf
u}\nabla {\mathbf u} = -\nabla p +\frac{1}{Re} \nabla^2 {\mathbf u}
\end{equation} 

\begin{equation}\label{eq:NS2} 
\nabla \cdot {\mathbf u} = 0.
\end{equation}  

\begin{equation}\label{eq:Energy} 
\frac{\partial T}{\partial t}+\mathbf u
\nabla T= \frac{1}{Re Pr} \nabla^2 T 
\end{equation} 

\begin{figure}
\includegraphics[width=0.88\textwidth]{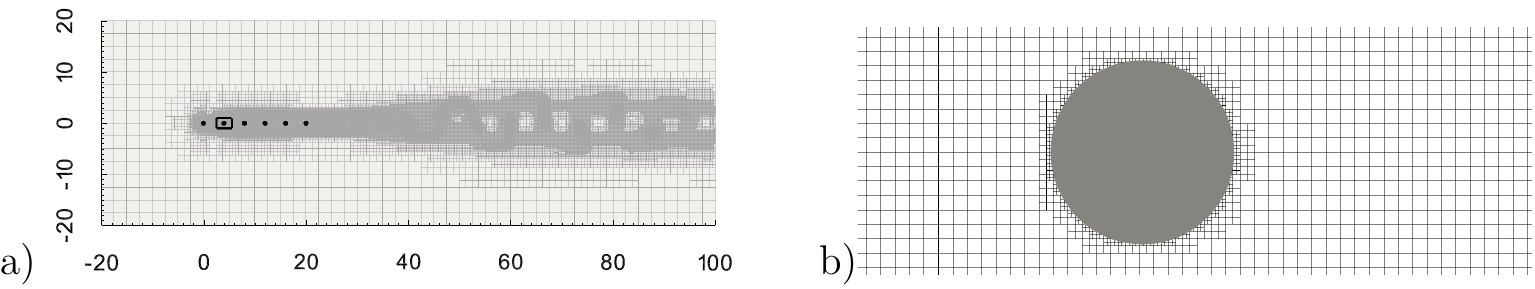} \caption{a) Instantaneous
grid resolution (see  grid $2$ in table \ref{tab:grid}) of six in-line
circular cylinders at a spacing ratio of $4$.  b) The zoomed view of the grid
around the second cylinder highlighted with a square box in
a).}\label{fig:grid-sketch} 
\end{figure}

The stream-wise and transverse directions of the Cartesian reference frame are
defined as $x$ and $y$, respectively.   The dimensionless velocity vector is
${\mathbf u} = (u, v)$ with $u$ and $v$ the stream-wise and the transverse
velocity components, respectively. The dimensionless pressure, temperature and
time are, respectively, $p$, $T$ and $t$. Direct numerical simulations has
been considered by mean of Gerris numerical code described in
~\cite{popinet2003} and ~\cite{popinet2009}. The equations (\ref{eq:NS1}),
(\ref{eq:NS2}) are solved using a classical fractional step projection method.
The advection terms in the equations (\ref{eq:NS1}) and (\ref{eq:Energy}) are
computed using a Godunov procedures with a second order upwind method and  the
viscous terms are treated implicitly with a Crank-Nicholson method. The
cylinder surfaces within the domain are treated with a volume-of-fluid
approach (\cite{popinet2009}). The boundary conditions at the inflow are
defined Dirichlet for the velocity and the temperature with constant values $u
= 1$, $v = 0$ and $T=0$ and Neumann for the pressure, $\partial{p}/\partial  x
= 0$.  At the outflow there are Dirichlet boundary condition for the pressure
$p = 0$ and Neumann condition for the velocity components and the temperature
$\partial{(u,v,T)}/\partial x = 0$. In order to avoid any influence of the
outflow on the dynamics of the shedding vortices, the outflow boundary has
been placed $60$-$80$ cylinder diameters far from the downstream cylinder,
where the wake disturbances have been dissipated due to the viscosity
(\cite{Kloker1993620}).  The boundary conditions in the transverse direction
is set as free-slip, $\partial{u}/\partial{y}=0$, $v=0$ and $\partial
T/\partial y=0$. A Dirichlet boundary condition for the temperature is set at
the solid cylinder surfaces, $T=1$.  An adaptive mesh refinement with a
quad-tree approach is used. A static mesh refinement at the solid surfaces and
a dynamical mesh refinement generated according to the local velocity and
temperature gradient are implemented. In figure \ref{fig:grid-sketch} a sketch
of the whole numerical domain for the six in-line cylinders configuration and
an instantaneous cells distribution behind the downstream cylinder is
depicted. In table \ref{tab:grid} the details of the computational grid are
listed.  The total amount of grid points cannot be reported since they are
time dependent.  The code is parallelized using distributed memory approach
(MPI protocol) so multi-processors simulations have been performed. The
simulations has been run on a Linux cluster.  More details on the numerical
code (Gerris) are described in \cite{popinet2003}.  

\section{Validation study} 

\subsection{Flow around a single cylinder} According to the well known problem
of a single cylinder immersed in a stream flow, a validation study about the
computational grid has been performed at $Re=100$. The dimensions of the
numerical domain have been chosen in order to reduce the boundary influence on
the solution. The boundaries distances from the center of the cylinder are:
$20d$ from the inlet, $20d$ from the lower and upper boundaries and $60d$ from
the outlet. Four different grid resolutions have been chosen: grid $1$,  grid
$2$, grid $3$ and  grid $4$ from the coarsest to the finest, respectively. The
grid details are reported in table \ref{tab:grid}.  No artificial perturbations
are added to the boundary condition in order to trigger the vortex shedding.
For the above listed tests we get the stationary solution after $150$
dimensionless time units, after that we collect the statistics for $250$ time
units.

%\begin{table} \centering
%\begin{tabular}{ c p{1.8cm} p{1.8cm} p{1.8cm} p{1.4cm} p{1.8cm} c }
%{Case} & Cylinder grid points & Minumum grid size & Maximum grid size &
%$\overline{C_d}$, ~~~~~~~~~ $\sigma$ & $\overline{C_d}$ variation $(\%)$ &
%$St$ \\\\ 
%1 &  $84$ & $0.0391 d$ & $2.50 d$  & $1.3718$,
%$0.00744 $ & $2.2$ &  $0.1636$\\ 
%2 & $148$ & $0.0195 d$ & $2.50 d$  & $1.3488$,
%$0.00699 $ & $0.49$ & $0.1636$\\ 
%3 & $352$ & $0.00977 d$ & $2.50 d$  & $1.3427$,
%$0.00671 $ & $0.037$ & $0.1636$\\
%4 & $352$ & $0.00977 d$ & $0.625 d$ & $1.3422$, 
%$0.00670 $ & $0$ & $0.1636$\\
%\end{tabular} 

\begin{table*} \centering
%\begin{tabular}{ c p{1.8cm} p{1.8cm} p{1.8cm} }
\caption{Grid resolution details. The cylinder grid points represent the
number of cells along the solid surface. The minimum and the maximum grid
sizes are computed in the whole domain. The total number of cells is not
reported since it changes in time.}\label{tab:grid} 
%\begin{ruledtabular}
\begin{tabular}{ c c c c }
%\hline
{Grid} & Cylinder grid points & Minimum grid size & Maximum grid size \\ 
\hline
1 &  $84$ & $0.0391 d$ & $2.50 d$   \\ 
2 & $148$ & $0.0195 d$ & $2.50 d$   \\ 
3 & $352$ & $0.00977 d$ & $2.50 d$  \\
4 & $352$ & $0.00977 d$ & $0.625 d$ \\
%\hline
\end{tabular} 
%\end{ruledtabular}
\end{table*}

\begin{table*} \centering
\caption{ Quantitative comparison of the results at different grid resolutions
with those in literature at $Re = 100$ for single
cylinder configuration. The standard deviation of the drag and lift coefficients 
($\sigma(C_D)$, $\sigma(C_L)$) is directly calculated on the time series of 
$C_D$ and $C_L$ for \cite{FLD:FLD812} and the present work. For the other authors a sine wave
fluctuation has been assumed for the fluctuating coefficients and their pick
amplitudes are divided by $\sqrt{2}$.}\label{tab:singlecylinder}    
%\begin{ruledtabular}
%\begin{tabular}{ p{1.0cm}  p{2.0cm} p{2.0cm} p{2.3cm} c c c c }
\begin{tabular}{ c c c c c }
%\hline
& $C_D$ & $\sigma(C_D)$ & $\sigma(C_L)$ & $St$\\
\hline
\cite{FLM:8322063}& $1.338 \pm 0.009$ & $0.0064$ & $0.23$ & $0.164$ \\
\cite{Park}& $1.33 \pm 0.009$ & $0.0064$ & $0.23$ & $0.165$ \\
\cite{FLD:FLD812} & $1.33 \pm 0.009$ & $0.0064$ & $0.23$ & $0.164$ \\ 
\cite{Ding2007}&  $1.356 \pm 0.0010$ & $0.0071$ & $0.20$ & $0.166$ \\ 
grid 1 &  $1.371 \pm 0.0105$ & $0.0074$ & $0.23$ & $0.161$ \\ 
grid 2 &  $1.349 \pm 0.0098$ & $0.0070$ & $0.23$ & $0.165$ \\ 
grid 3 &  $1.343 \pm 0.0095$ & $0.0067$ & $0.23$ & $0.165$ \\ 
grid 4 &  $1.342 \pm 0.0095$ & $0.0067$ & $0.23$ & $0.165$ \\  
%\hline
%\end{ruledtabular}
\end{tabular} 
\end{table*}

%\begin{figure} \fontsize{14}{16} \centering \scalebox{0.48}{\input{cl-fft.tex}}
%\normalsize \caption{FFT of the lift force coefficient $C_L$ at different grid
%resolutions. The amplitude is normalized
%respect to the maximum. (Colour online).}\label{fig:fft-test} \end{figure}

The control parameters are: the time-average
drag coefficient, $\overline{C_D}$, the standard deviation of the drag
and the lift coefficients, $\sigma(C_D)$, $\sigma(C_L)$  and the Strouhal
number, $St=f d/U_0$ where $f$ is the frequency of $C_L$. 
%In figure \ref{fig:fft-test} the Fourier
%spectra of the lift coefficient $C_L$ have been reported for all the test
%cases that represents the non dimensional wake shedding frequency.  
The results of the test cases are compared with the results of \cite{FLM:8322063},  \cite{Park}, \cite{FLD:FLD812}, 
\cite{Ding2007}(see table \ref{tab:singlecylinder}).  The
results are in agreement with the data in literature. Thus, 
the grid resolution of  grid $2$ is the best
compromise between computational cost and accuracy.  
%\begin{figure} \fontsize{22}{24} \centering
%\scalebox{1.0}{\input{cylinder-drag.tex}} \normalsize \caption{Comparison of
%the measured data of the average drag coefficient $\overline{C_D}$ ($\Box$)
%with the literature data of \cite{panton1984}(+)}\label{fig:cylinder-drag}
%\end{figure}

%\begin{figure} \fontsize{14}{16} \centering \scalebox{0.48}{\input{cl-fft-two.tex}}
%\normalsize \caption{FFT of the lift force coefficient of the downstream
%cylinder ($C_L^{(2)}$) 
%for two tandem cylinder configuration. The amplitude is normalized
%respect to the maximum. (Colour
%online).}\label{fig:vort-fft-two} \end{figure}

\subsection{Flow around two tandem cylinders} 

As described in the introduction the flow around multiple bodies implies a
flow field that could be completely different compared to the flow around a
single body. Especially in the in-line configuration the upstream body
influences the flow dynamics on the downstream ones.  In \cite{FLD:FLD812} the
fluid flow around two circular cylinders in a tandem configuration has been
described. The force characteristics on both the cylinders are excerpted. In
the present work numerical experiments at a spacing ratio of $3.5$, $3.75$,
$4$ and $6$ have been performed.  All the tests converge in $400$
dimensionless time units except the case at a spacing ratio of $4$ that
converges after about $800$ time units.  All the statistics have been
collected for $1000$ time units after the convergence.  For these simulations
the dimensions of the numerical domain are $120d\times40d$ in the stream-wise
and transverse direction, respectively. The boundary distances from the center
of the upstream cylinder are: $20d$ from the inlet, the lower and the upper
boundary and $100d$ from the outlet.
\begin{figure}[!ht] 
%\fontsize{12}{14} \centering 
%\scalebox{0.48}{{\huge a)}\input{cd-Sharman.tex}}
%\scalebox{0.48}{{\huge b)}\input{cdrms-Sharman.tex}}\\
%\scalebox{0.48}{{\huge c)}\input{clrms-Sharman.tex}}
%\scalebox{0.48}{{\huge d)}\input{st-Sharman.tex}}
%\normalsize 
\includegraphics[width=0.88\textwidth]{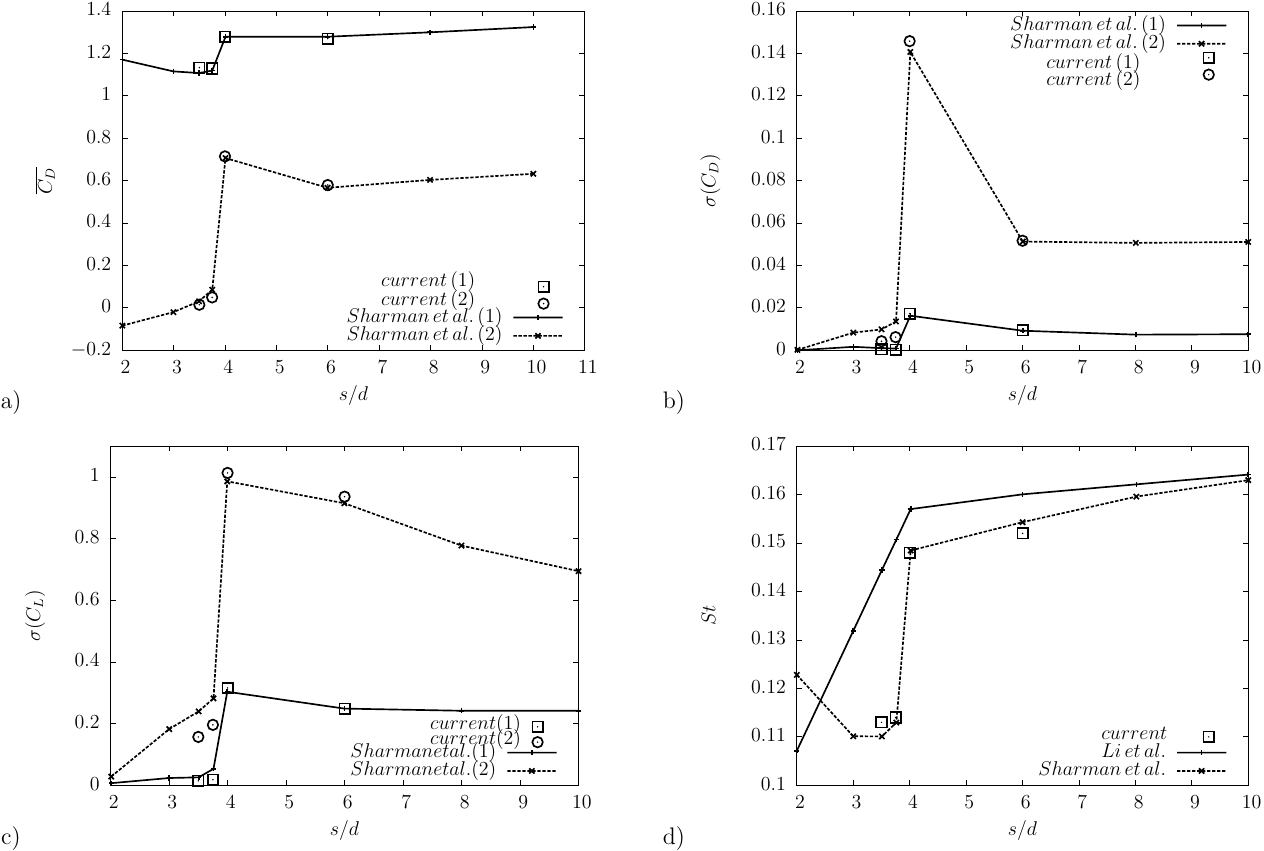}
\caption{Comparison of the drag coefficient $C_D$ (a), standard
deviation of the drag coefficient $\sigma(C_D)$ (b), standard deviation of the
lift coefficient $\sigma(C_L)$ (c) and no-dimensional shedding frequency $St$
(d) for upstream (1) and downstream cylinder (2) for two tandem cylinders
respect to the results of \cite{FLD:FLD812} and \cite{Li1991155}.}
\label{fig:Sharman} \end{figure}
In fig. \ref{fig:Sharman}a and \ref{fig:Sharman}b the comparison of the time
average, $\overline{C_D}$, and the standard deviation, $\sigma(C_D)$, of the
drag coefficients for the upstream (1) and downstream cylinders (2), respect
to \cite{FLD:FLD812},  are reported. The distance for which the mean drag
coefficient of the second cylinder is significantly enhanced is called
critical center-to-center spacing. It is correctly reproduced by the current
test as reported in fig. \ref{fig:Sharman}a.  
\begin{figure}[!ht]
\includegraphics[width=0.88\textwidth]{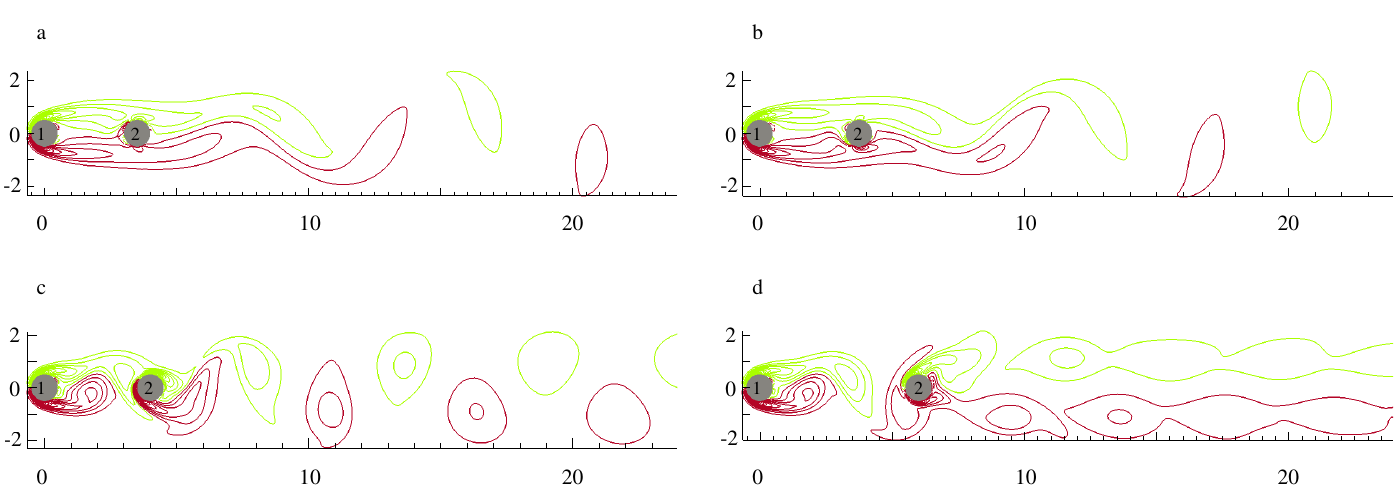}
\caption{Instantaneous vorticity iso-lines for a spacing of $3.5 d$ (a),
$3.75d$ (b), $4 d$ (c) and $6 d$ (d) in two tandem circular cylinders
configuration. Green (light gray) lines for negative values, red (dark grey)
for positive values. (Colour online).}\label{fig:vort-iso-two} 
\end{figure}
The vorticity patterns at different spacing are reported in figure
\ref{fig:vort-iso-two} where the flow pattern transition between a spacing
ratio of $3.75$ and $4$ is shown. In figures \ref{fig:vort-iso-two}a-b for a
spacing ratio of $3.5$ and $3.75$ respectively the vorticity field in the gap
appears split in an upper negative vorticity zone (green/light grey) and a
lower positive vorticity zone (red/dark grey).  The wakes are elongated in the
stream-wise direction.  For a spacing ratio of $4$ and $6$ (see figure
\ref{fig:vort-iso-two}c-d) the wake shedding is completely triggered and it
influences the gap region. The unsteady wake of the first cylinder interacts
with the cylinder $2$ generating a more unstable wake downstream.  The
differences in the shedding frequency, for changing the spacing ratio, can be
seen qualitatively looking at the vorticity iso-lines of the wakes (figure
\ref{fig:vort-iso-two}) and quantitatively in figure \ref{fig:Sharman}d with
the results of \cite{FLD:FLD812} and \cite{Li1991155}. As reported in
\cite{FLD:FLD812} the results of \cite{Li1991155} are not reliable according
to insufficient domain size, grid resolution and too large time step. Whereas,
there is a good agreement with the results of \cite{FLD:FLD812}.

\section{Results: six in-line circular cylinders} 
\subsection{Flow characteristics} 

The flow around six in-line cylinders is investigated. For this configuration
\cite{Liang2009950} found
a transition in the values of the average drag coefficient at $Re=100$ between
a spacing ratio of $3.6 - 4$. 
\begin{figure}[!ht]
\includegraphics[width=0.88\textwidth]{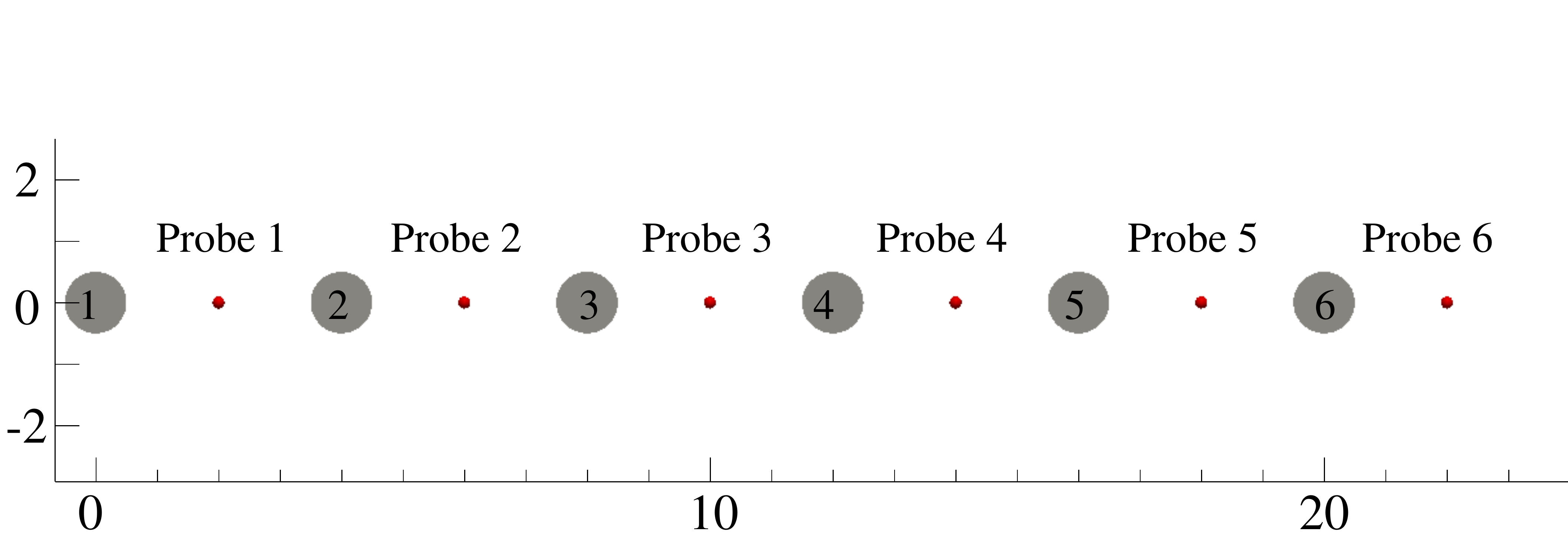} \caption{Sketch of the
six probe positions within the numerical domain.}\label{fig:probe-position} 
\end{figure} 
One probe per each gap has been located between the cylinders and one behind
the last cylinder, where the pressure, the velocity components and the
 vorticity are sampled (see figure \ref{fig:probe-position}). The
statistics are collected from $1000$ to $1800$ dimensionless time units in
order to ensure the convergence with a sufficient number of samples.

The results for six in-line circular cylinders have no reference data in
literature, except the numerical results of \cite{Liang2009950}. In table
\ref{tab:LiangComparison} the main Strouhal number of the vortex shedding for
$s/d=3.6$ and $s/d=4.0$ are compared. Some differences are present, likely
because of some lack in the domain size and statistical accuracy of
\cite{Liang2009950} results. The Liang outflow has been set $32d$ far from the
last cylinder respect to $80d$ considered in the present results. The
following flow patterns analysis will highlight that the far field wake
evolution could affect the equations solution, especially for the case of six
in-line cylinder configuration. Moreover, here about $100$ shedding periods of
flow transient phase have been discarded and then the statistics have been
collected for $180$ shedding cycles. Otherwise, the $20$ transient shedding
periods and $10$ shedding cycles of statistics considered by Liang with the
same initial conditions (zero velocity and pressure field) are not enough to
ensure the statistical convergence of the mean quantities, according to the
long transient phase that this configuration has, and the few statistical
samples considered.

\begin{table*} \centering
%\begin{tabular}{ c p{1.8cm} p{1.8cm} p{1.8cm} }
\caption{Comparison of the main Strouhal number of the vortex shedding between
\cite{Liang2009950} and the present results }\label{tab:LiangComparison} 
%\begin{ruledtabular}
\begin{tabular}{ c c c }
%\hline
$s/d$ & \cite{Liang2009950} & Present results \\ 
\hline
$3.6$ &   $0.109$  & $0.106$   \\ 
$4.0$   &   $0.1502$ & $0.138$     \\ 
%\hline
\end{tabular} 
%\end{ruledtabular}
\end{table*}

\begin{figure}
\includegraphics[width=0.88\textwidth]{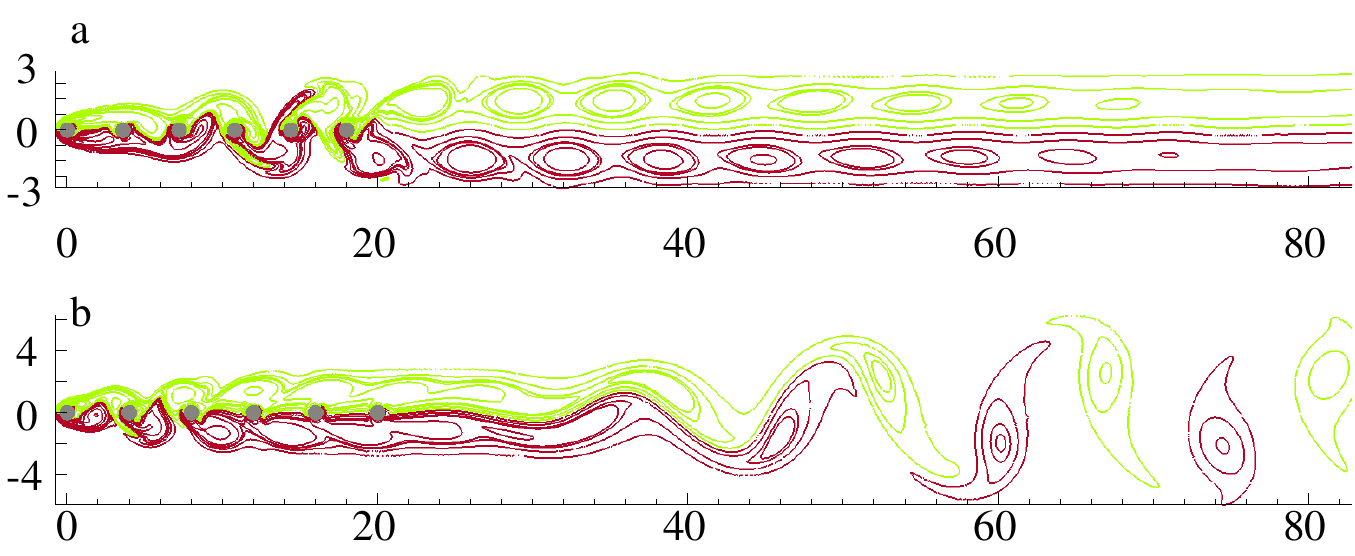}
\caption{Far field instantaneous vorticity iso-lines for a spacing of $3.6 d$
(a) and $4 d$ (b). Green (light gray) lines for negative values, red (dark
grey) for positive values in six in-line circular cylinders configuration.
(Colour online).}\label{fig:farfield} \end{figure} 
%\begin{figure}[!ht]
%\includegraphics[width=0.88\textwidth]{vort-isolines-1x6}
%\caption{Instantaneous vorticity iso-lines for a spacing of $3.6 d$ (a) 
%and $4 d$ (b). Green (light gray) lines for negative values, red (dark grey)
%for positive values in six in-line circular cylinders configuration. (Colour
%online).}\label{fig:vort-3.6-4}
%\end{figure}
In figure \ref{fig:farfield} the instantaneous vorticity contours for a
spacing ratio equal to $3.6$ and $4$  are shown. For a $3.6$ spacing ratio
(figure \ref{fig:farfield}a) there is a quasi-steady recirculation flow in the gap
between the cylinders $1$ and $2$. On the other hand, starting from cylinder
$2$ to cylinder $6$, the vortex shedding can be distinguished. 
\begin{figure}[!ht]
\includegraphics[width=0.88\textwidth]{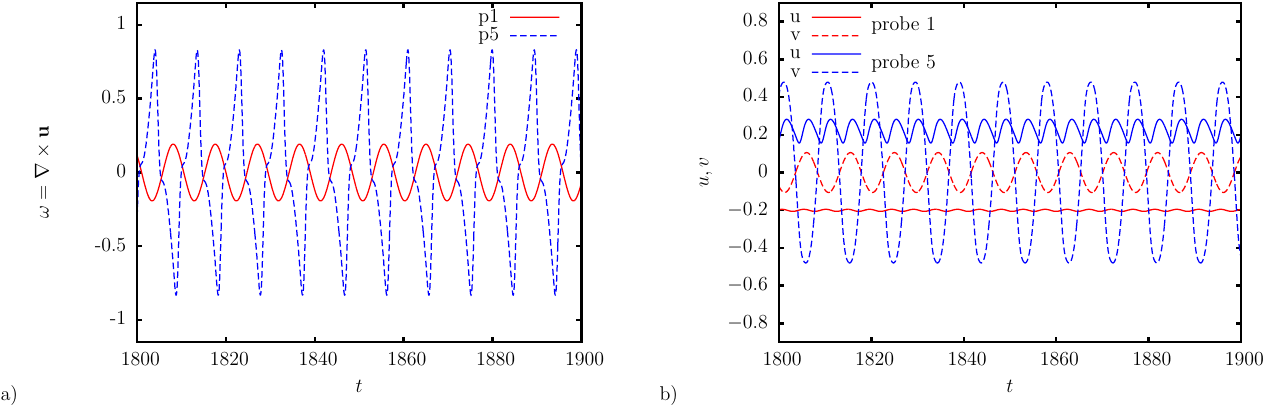} 
\caption{Six in-line cylinders configuration for a spacing of $3.6d$.
a) Time series of  vorticity, $\omega_z$, b) stream-wise velocity
component (solid line), $u$,
and transverse velocity component (dashed line), $v$. The data are collected in the middle
of the gap between the cylinders $1$ and $2$ (red), probe $1$, and between the cylinders
$5$ and $6$ (blue), probe $5$.
(Colour online).}\label{fig:vort-vel-3.6} \end{figure}

In figure \ref{fig:vort-vel-3.6} the time series of the vorticity and of the
velocity components collected by the probe $1$ and the probe $5$ are shown.
The vorticity time series in the gap between cylinder $1$ and $2$ (probe $1$)
has a sinusoidal behaviour with an amplitude smaller than those recorded in
the gap between cylinder $5$ and $6$ (probe $5$) where the vortex shedding
strongly affects the flow field (see figure \ref{fig:vort-vel-3.6}a).
Therefore, the negative stream-wise velocity component in probe $1$ confirms
the presence of a quasi steady recirculation flow between cylinder $1$ and $2$
(see figure \ref{fig:vort-vel-3.6}b).   The force components have been
computed for each cylinder as dimensionless force coefficients: lift , $C_L$,
and drag $C_D$.  \begin{figure}[!ht] \centering
\includegraphics[width=0.88\textwidth]{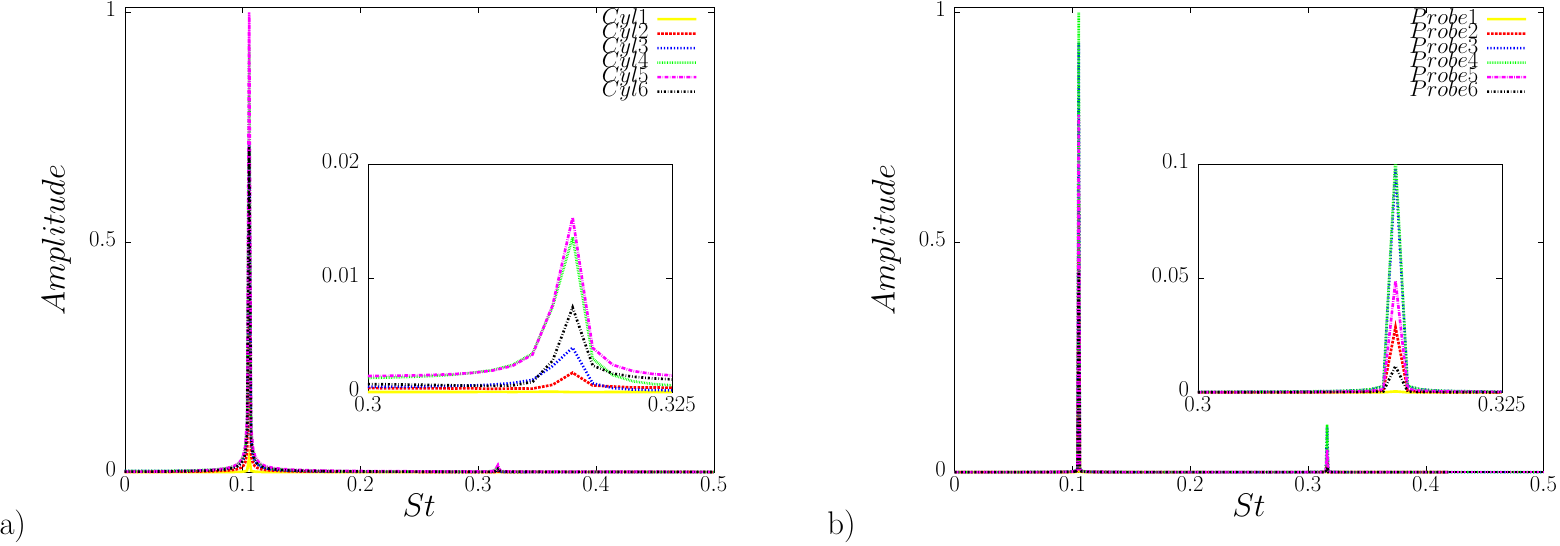} \caption{a)
FFT of the lift force coefficients at a spacing ratio of $3.6$ for six in-line
cylinders configuration. b) FFT of the transverse velocity component at a
spacing ratio of $3.6$ for six in-line cylinders configuration. The amplitude
is normalized respect to the maximum. (Colour online).}\label{fig:fft-six-3.6}
\end{figure} 

In figure \ref{fig:fft-six-3.6}a the amplitude spectra of the $C_L$ for a
spacing ratio of $3.6$ is shown. For all the cylinders the main shedding
frequency is  equal to $0.106$ .  A high secondary frequency at $St=0.316$,
equal to the third harmonic of the above mentioned main frequency, can be
distinguished on cylinder $5$ with very narrow band-width. These small
structures, coming from upstream region (probe $4$), are dissipated behind the
last cylinder (probe $6$).  Looking at figure \ref{fig:farfield}a,  the case
at $3.6$ spacing ratio reveals, in the wake of the last cylinder, two rows of
counter-rotating shedding vortices. We call this mode: stable shear layer
(SSL) mode.  

%Compared with the case of two
%cylinders (\cite{PhysRevE.81.036305}), the flow field is similar to the so-called
%secondary vortex formation (SVF) mode, where the more stable shear layers of
%the six cylinders configuration  avoids the formation of a secondary street,
%leaving the viscosity to dissipate the wake structures.  

\begin{figure}
\includegraphics[width=0.88\textwidth]{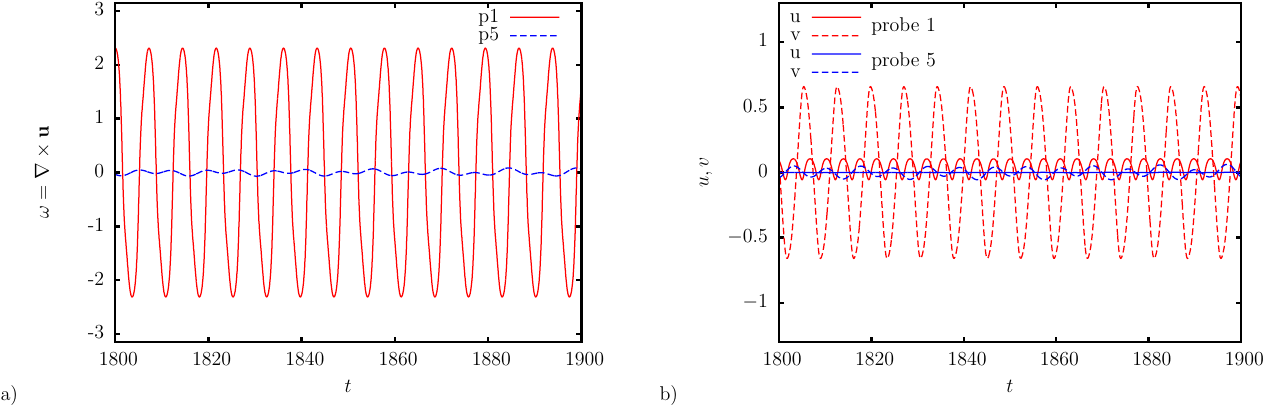} 
\caption{Six in-line cylinders configuration for a spacing of $4d$.
a) Time series of  vorticity, $\omega_z$, b) stream-wise velocity
component (solid line), $u$,
and transverse velocity component (dashed line), $v$. The data are collected in the middle
of the gap between the cylinders $1$ and $2$ (red), probe $1$, and between the cylinders
$5$ and $6$ (blue), probe $5$.
(Colour online).}\label{fig:vort-vel-4} \end{figure}

\begin{figure} 
\centering
\includegraphics[width=0.88\textwidth]{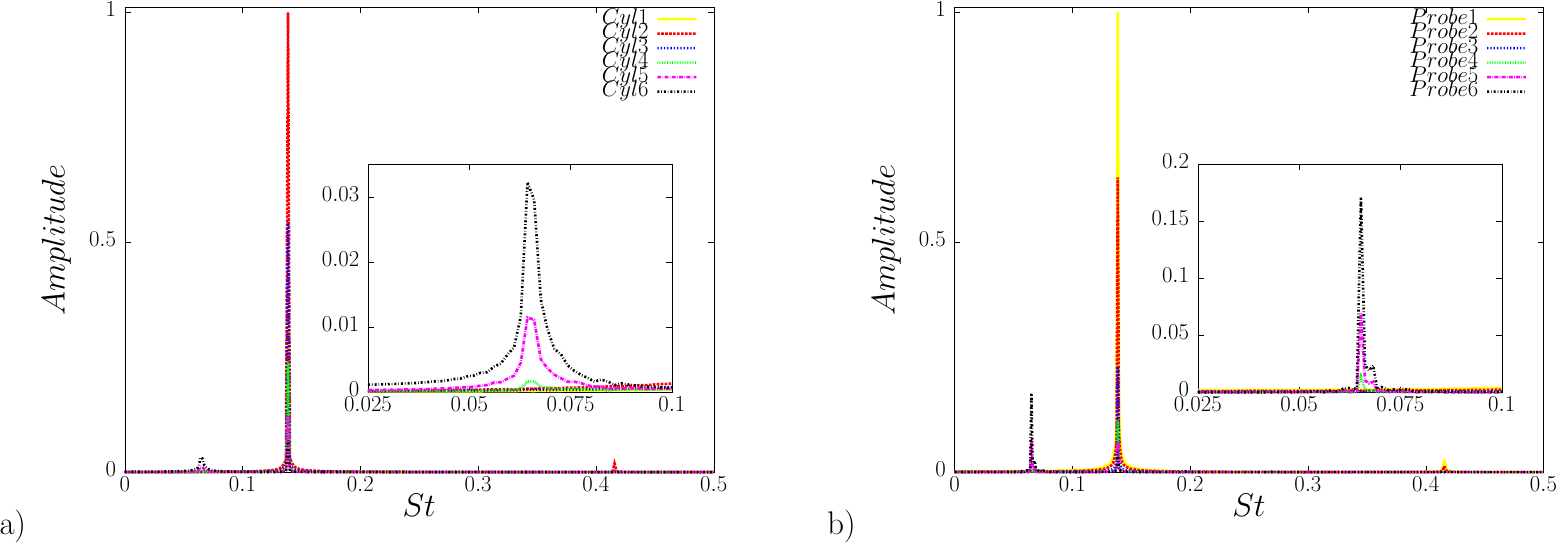} 
\caption{a) FFT of the lift
force coefficients at a spacing ratio of $4$ for six in-line cylinders
configuration. b) FFT of the transverse velocity component at a spacing
ratio of $4$ for six in-line cylinders configuration. The amplitude is normalized respect to the maximum. (Colour
online).}\label{fig:fft-six-4} \end{figure} 

For a spacing ratio of $4.0$ different flow patterns can be distinguished
respect to the case at $3.6$ (see figure \ref{fig:farfield}b).  The vortex
shedding takes place at the upstream cylinders ($1$ - $2$) and it interacts
with cylinder $3$ producing two opposite vorticity shears that slip along the
successive cylinders. In the gap regions of the downstream cylinders there are
low vorticity zone as showed in fig\ref{fig:vort-vel-4}, where the amplitude
of the vorticity in the probe $5$ is smaller than that in the probe $1$. The
amplitude spectra of the $C_L$ in the cylinder positions and of transverse
velocity component, $v$, in the probe positions for $s/d=4$ are reported in
fig.\ref{fig:fft-six-4}. The main oscillating frequency is increased from
$St=0.106$ at $s/d=3.6$ to $St=0.138$ at $s/d=4.0$. Varying the spacing from
$3.6d$ to $4d$ plays an important role in the modification of the flow
structures leading to a change of the heat exchange as described in detail in
the following section. In this case too,  there is an high frequency peak,
$St=0.416$, it is equal to the third harmonic of the main frequency,
$St=0.138$.  Moreover at $s/d=4.0$ there is the presence of the low frequency
peak in the amplitude spectra of $C_L$ and $v$, see figure
\ref{fig:fft-six-4}.  This low frequency peak is generated by the flow field
instability in the wake region. The vorticity pattern in the wake region (see
fig.\ref{fig:farfield}b) highlights that the two shear layers take place at
the cylinder $3$ and the cylinder $4$ and they slip along the downstream
cylinders. After about $10d$ downstream the last cylinder ($6$) the shear
layers become unstable and a secondary wake starts to shed vorticity at a low
frequency respect to the shedding frequency of the upstream cylinders. We call
this mode shear layer secondary vortices (SLSV) mode. 

A quantitative evidence of the flow transition between SSL and SLSV mode is
showed in table \ref{tab:singlerow}. The value of $\overline{C_D^{(2)}}$
suddenly increases from $-0.0159$ to $0.4212$ at a spacing ratio of $3.6$ and
$4$, respectively.  On the other hand there is a sudden decrease of the
each time-average drag coefficient of the cylinders $4$, $5$, $6$
($\overline{C_D^4}$, $\overline{C_D^5}$, $\overline{ C_D^6}$) for increasing the
spacing ratio from $3.6$ to $4$. It is worth to note the drag reduction on the
entire row. The overall mean drag coefficient ($\overline{C_D^{tot}}$)
decreases from $3.0$ to $2.1$ therefore the drag is reduced up to $30\%$.

\begin{table*} \centering

\caption{Time-average drag coefficients ($\overline{C_D}$) and standard
deviation ($\sigma$) for each cylinder. In bold, the negative value of the
drag coefficient for the second cylinder, $\overline{C_D^{(2)}}$, at
$s/d=3.6$, related to the {\it drag inversion
separation}}\label{tab:singlerow}

%\begin{ruledtabular}
\begin{tabular}{ c p{1.2cm} p{1.6cm} p{1.2cm} p{1.2cm} p{1.2cm} p{1.2cm} p{1.2cm} }
%\hline
$s/d$ & $\overline{C_D^{(1)}}$, ~~~~~~~~~ $\sigma$ & $\overline{C_D^{(2)}}$,  ~~~~~~~~~ $\sigma$ &  $\overline{C_D^{(3)}}$,  ~~~~~~~~~ $\sigma$ &
$\overline{C_D^{(4)}}$,  ~~~~~~~~~ $\sigma$ & $\overline{C_D^{(5)}}$,  ~~~~~~~~~ $\sigma$ &
$\overline{C_D^{(6)}}$,  ~~~~~~~~~ $\sigma$ & $\overline{C_D^{tot}}$,  ~~~~~~~~~ $\sigma$ \\
\hline
$3.6$ & $1.0872$, $0.0012$ & $\mathbf{-0.0159}$, $0.0127$ & $0.5380$, $0.0805$ &
$0.6606$, $0.1431$ & $0.5139$, $0.1309$ & $0.2528$, $0.0483$ & $3.0366$,
$0.1624$ \\ 
$4.0$ & $1.2158$, $0.0171$ & $0.4212$, $0.1137$ & $0.2191$, $0.0297$ &
$0.1069$, $0.0073$ & $0.0861$, $0.0016$ & $0.0991$, $0.0027$ & $2.1483$,
$0.1238$ \\
%\hline
\end{tabular} 
%\end{ruledtabular}
\end{table*}

\subsection{Heat transfer analysis} 

\begin{figure} 
\centering
\includegraphics[width=0.65\textwidth]{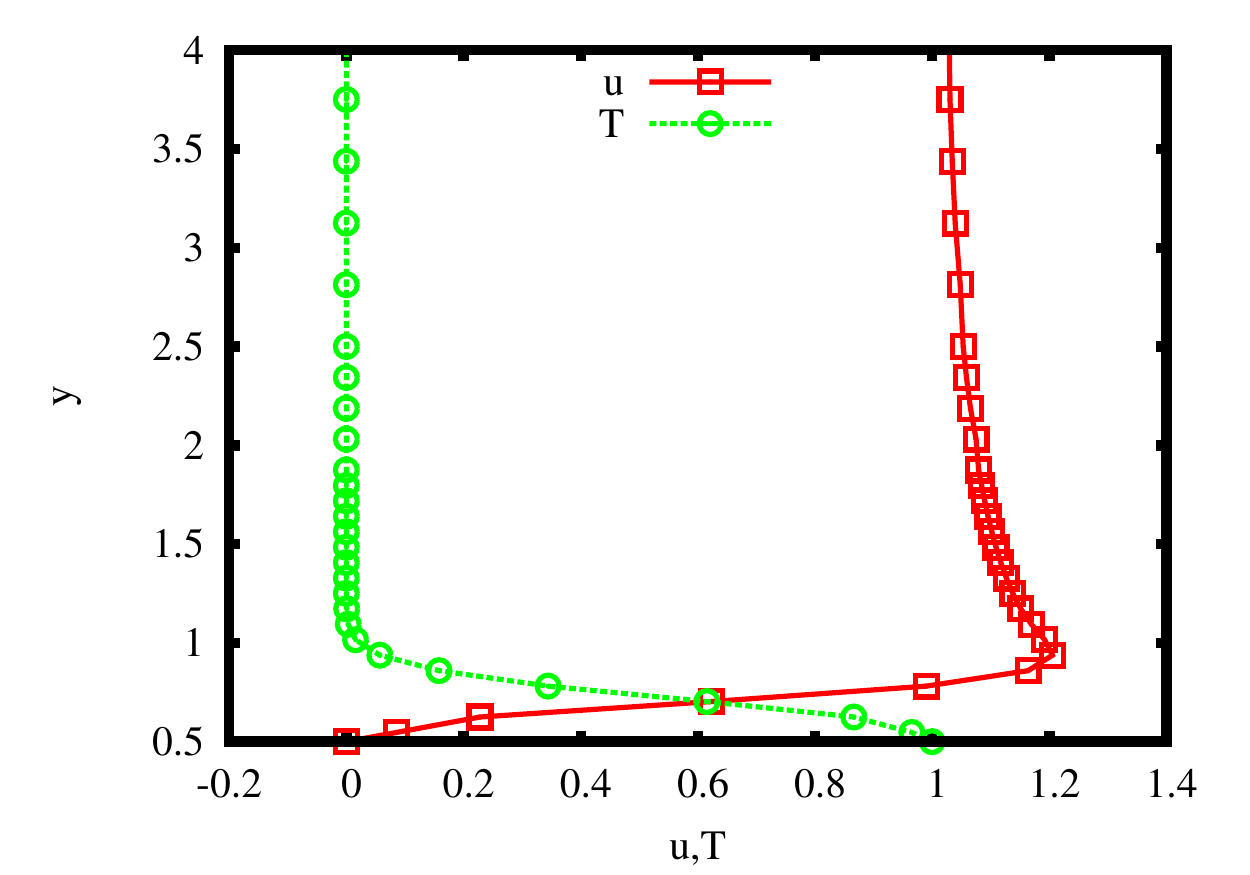} 
\caption{Instantaneous dimensionless velocity and temperature profiles above the
first cylinder for the case at $s/d=4$. $y$ is the dimensionless trasversal coordinate where $y=0.5$ corresponds to the cylinder surface (Colour
online).}\label{fig:bl} \end{figure}

\begin{figure}[ht!] \centering
\includegraphics[width=0.88\textwidth]{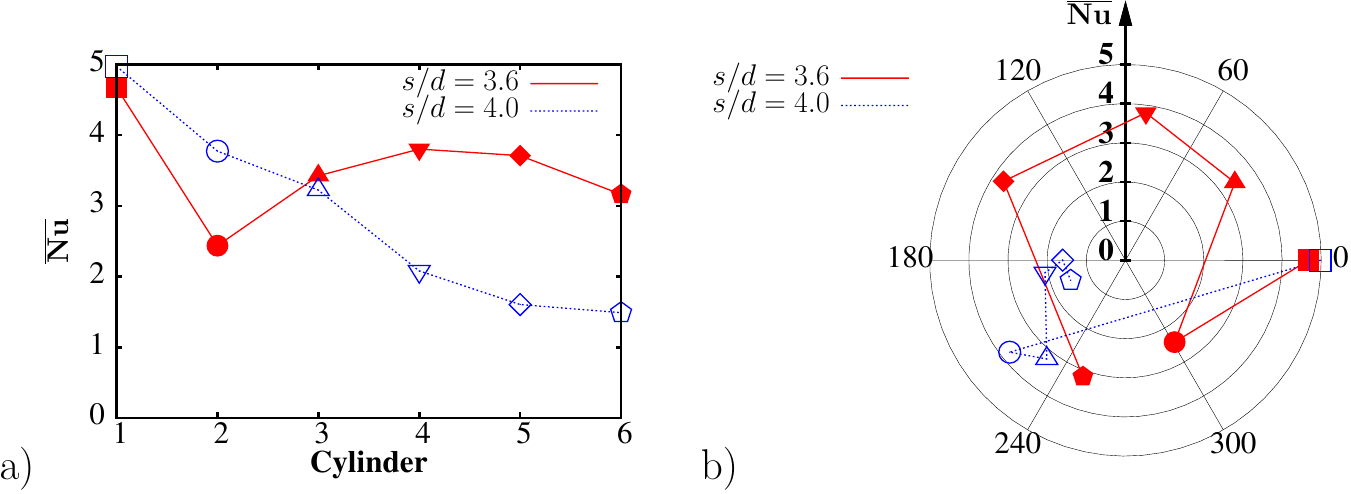} 
\caption{ a) The time-average mean Nusselt numbers of each cylinder for $s/d=3.6$ (filled symbols) and $s/d=4$ (open symbols);
b)  polar diagram of the time-average mean Nusselt numbers of the
cylinders with their phase shifts. Where the phase of
cylinder $1$ has been kept as reference, $0^\circ$. The cylinders are ordered
along the connecting line from cylinder $1$ (square symbol) to cylinder $6$
(pentagon symbol). (Colour online).}\label{fig:nu-split} \end{figure}

The analysis of the flow field around six in-line circular cylinders reveals
the transition between different patterns according to the space ratio, $s/d$.
In this section the ability of the flow structures in enhancing the heat
transfer between the cylinders and the surrounding fluid  has been
investigated for the spacing ratio of $3.6$ and $4$. A Prandtl number equal to
$0.7$ has been considered in order to reproduce the heat transfer for a fluid
with air thermal diffusivity.  For all the simulations the temperature of the
cylinders ($T_w^\ast$) and the temperature of the inflow ($T_\infty^\ast$) has
been kept constant.  The dimensionless temperature is defined as
$T=(T^\ast-T^\ast_\infty)/(T^\ast_w-T^\ast_\infty)$. Being the thermal
boundary layer thicker than the viscous one ($\delta_{\nu} \sim {\rm Pr}\,
\delta_T$) no grid modification has been required. In order to check the
correctness of the boundary layer resolution near the cylinders, in
fig.\ref{fig:bl} the instantaneous dimensionless velocity and temperature
profiles above the first cylinder for $s/d=4$ have been reported, where $y$,
$u$ and $T$ are the dimensionless transverse coordinate, stream-wise velocity
component and the temperature, respectively. As follows in
fig.\ref{fig:nu-split}a, the first cylinder has the highest average Nusselt
number, thus the thinnest boundary layer is expected to be there.  Within the
thermal and viscous boundary layer thickness, there are at least $10$ grid
points (see fig.\ref{fig:bl}).

The heat flux transferred from the solid body to the
flow can be calculated as the conductive heat exchange at the solid wall being
the fluid velocity at the wall equal to zero:

\begin{align}\label{eq:heat_cond}
	dQ=-k\frac{\partial T^\ast}{\partial n^\ast}\frac{d}{2} d\theta =
-k\frac{(T_w^\ast - T_\infty^\ast)\partial T}{d \partial n}\frac{d}{2} d\theta = 
- \frac{k(T_w^\ast - T_\infty^\ast)}{2}\frac{\partial T}{\partial n}d\theta
\end{align}

where $k$ is the thermal conductivity, $\theta$ is the polar angle respect to
the center of the cylinder,
$\partial T^\ast/\partial n^\ast$ and $\partial T/\partial n$ are the
dimensional and the dimensionless temperature gradient normal to the cylinder
surface, respectively.

It has to be equal to the convective heat flux per unit length transferred that
reads:

\begin{align}\label{eq:heat_conv}
	dQ=h_\theta(T^\ast_w-T^\ast_\infty)\frac{d}{2} d\theta
\end{align}
with $h_\theta$ the local convective heat transfer coefficient. Comparing eq.\ref{eq:heat_cond} and eq.\ref{eq:heat_conv}:

\begin{align} 
%h_\theta = \frac{-k\left.\frac{\partial T^\ast}{\partial n^\ast}\right|_w}{\Delta}
h_\theta = -\frac{k}{d}\frac{\partial T}{\partial n}
\end{align}

The local Nusselt number represents the dimensionless local heat flux:

\begin{align} 
	Nu_\theta = \frac{h_\theta d}{k} = \frac{\partial T}{\partial n}
\end{align}

The mean (surface-averaged) Nusselt number along the cylinder surface is
defined as follows:

\begin{equation}\label{eq:meanNu}
Nu=\frac{1}{2\pi}\int_{\theta=0}^{2\pi}Nu_\theta \, d\theta.
\end{equation} 

\begin{figure}[ht!] 
\centering
\includegraphics[width=0.88\textwidth]{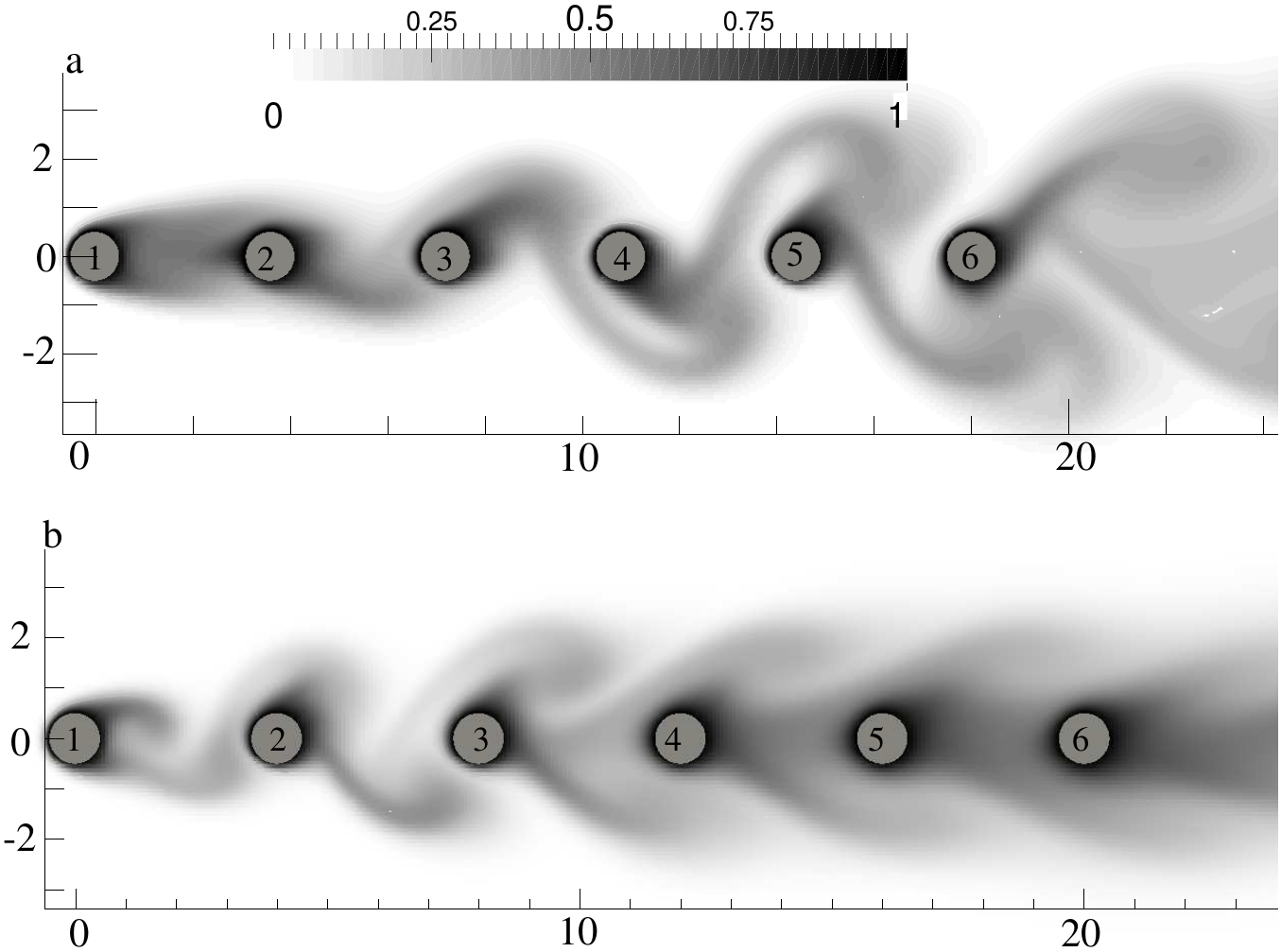} 
\caption{
Instantaneous temperature distribution around the cylinders for a
spacing ratio of $3.6$ (a) and $4$ (b) at $t=1800$.}\label{fig:temp-iso} 
\end{figure} 

\begin{figure}[ht!] 
\centering
\includegraphics[width=0.88\textwidth]{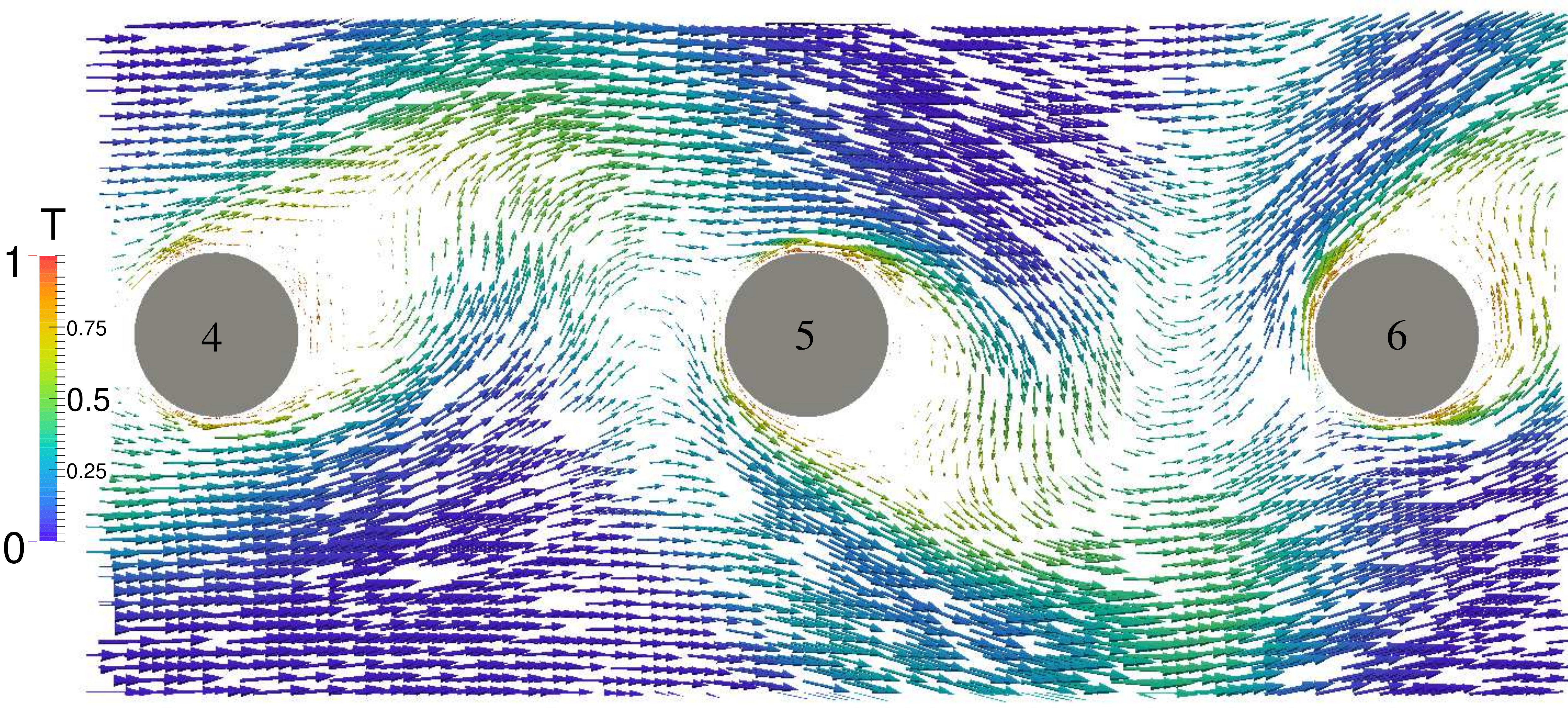} 
\caption{
Details of the instantaneous velocity vectors colored with the temperature
distribution around the cylinders $4$, $5$, $6$ for a
spacing ratio of $3.6$ at $t=1800$. (Colour online).}\label{fig:vect-temp} 
\end{figure} 

\begin{figure}[ht!] 
\centering
\includegraphics[width=0.65\textwidth]{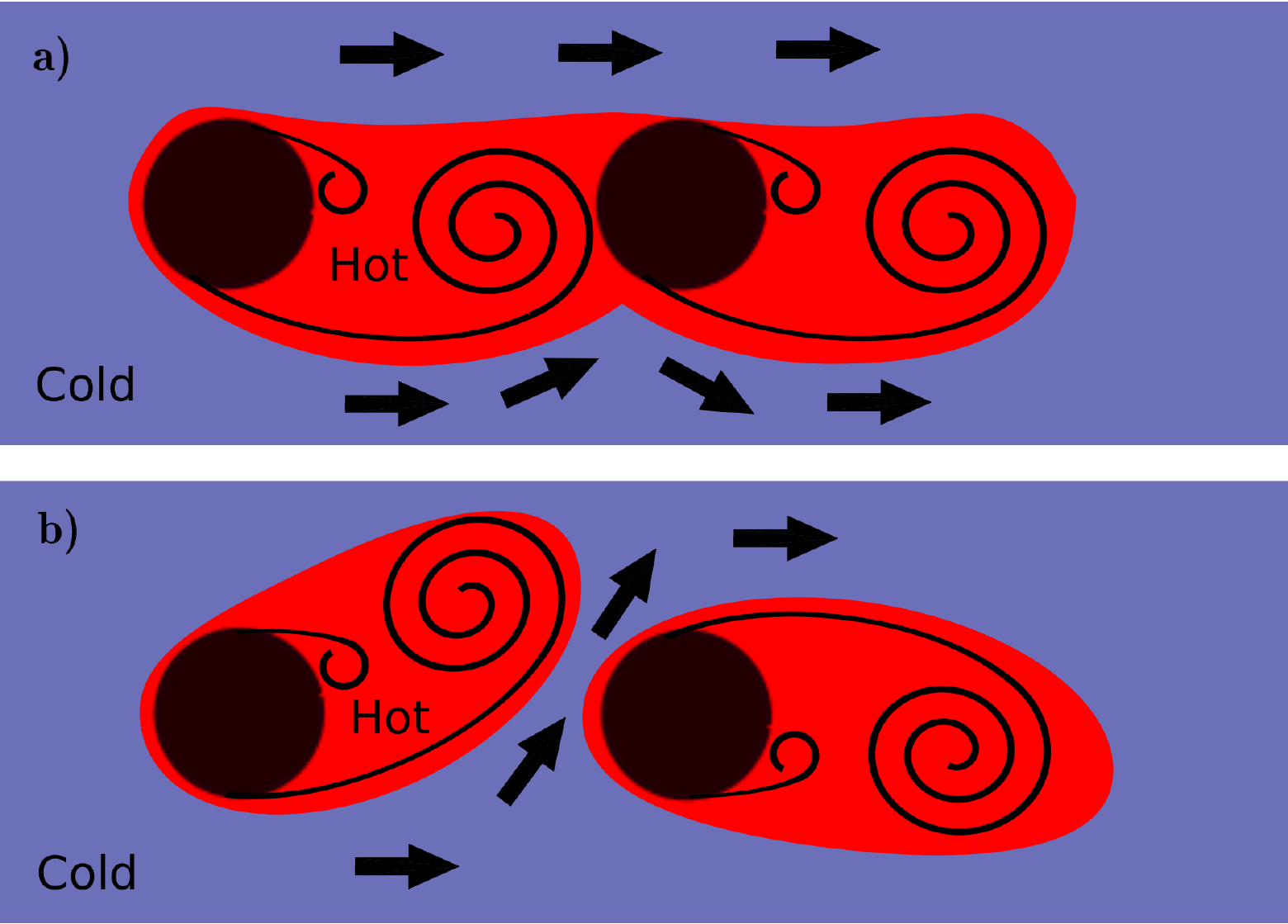}
\caption{Schematic of the cold fluid entrainment mechanism. a) in
phase vortex shedding (no entrainment), b) counter phase
vortex shedding (entrainment).}\label{fig:entrainment} 
\end{figure} 

\begin{figure}[ht!] 
\centering
\includegraphics[width=0.88\textwidth]{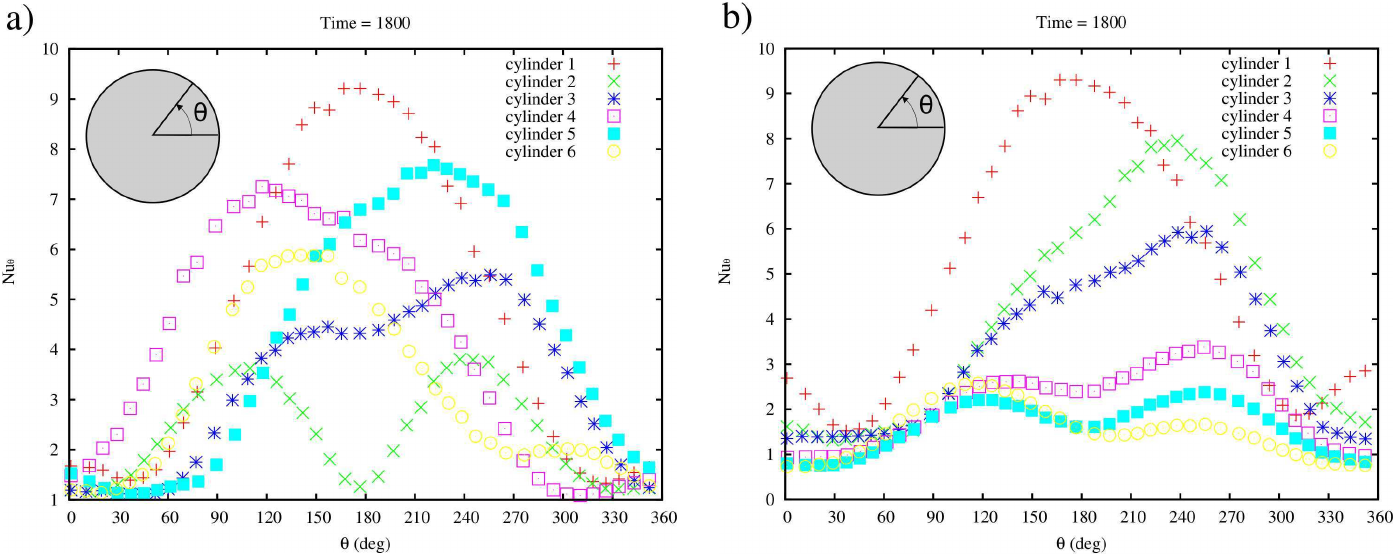} 
\caption{
Instantaneous local Nusselt number, $Nu_\theta$, distribution around the cylinders for a
spacing ratio of $3.6$ (a) and $4$ (b) at $t=1800$. (Colour online).}\label{fig:temp-distr} 
\end{figure} 

\begin{figure}[ht!] \centering
\includegraphics[width=0.88\textwidth]{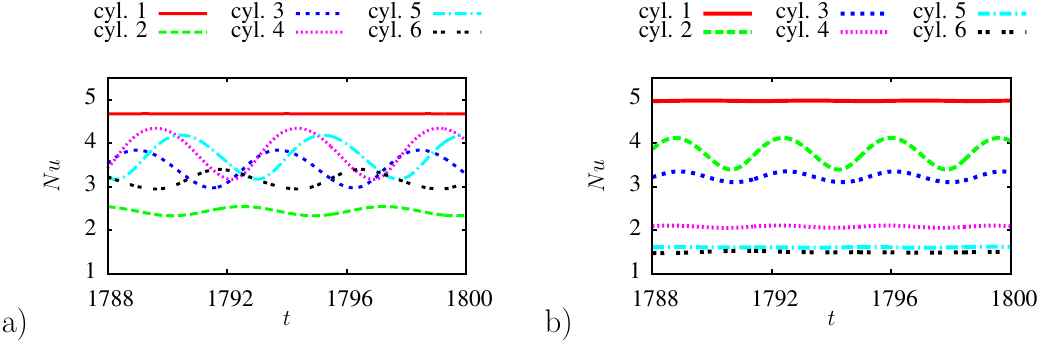}
\caption{Time series of the Nusselt number of each cylinder for $s/d=3.6$ (a)
and $s/d=4$ (b). (Colour online).}\label{fig:nu-ts} \end{figure} 

How the spacing between the cylinders affects the heat exchange?  The unsteady
behaviour of the flow field due to the oscillating vorticity behind the
cylinders affects the heat transfer.  The vorticity field influences the
temperature distribution of the fluid in the near field (see
fig.\ref{fig:temp-iso}).  In fig.\ref{fig:vect-temp} the velocity vectors
colored with their local temperature at $s/d=3.6$ shows that the flow field
causes the entrance of colder fluid over the cylinders. This phenomenon
enhances the heat exchange because of the higher thermal gradient between the
cylinders and the surrounding fluid. 

A schematic on how the flow pattern affect the heat exchange of the cylinders
is reported in fig.\ref{fig:entrainment}. The counter phase shedding causes
the entrainment of cold fluid in the gap region enhancing the heat exchange of
the downstream cylinder, on the contrary an in phase shedding avoid the
entrainment of cold fluid leaving the downstream cylinder immersed in a hot
region. At $s/d=4$ the downstream cylinders ($3-6$) shed vorticity in phase
otherwise at $s/d=3.6$ a counter phase shedding can be distinguished (see
fig.\ref{fig:farfield}). Otherwise, the cylinder $2$ has an in phase shedding
at $s/d=3.6$ and a counter phase shedding at $s/d=4.0$. Accordingly the
measured values of the Nusselt number for each cylinder
(fig.\ref{fig:nu-split}) confirms the relation between vorticity pattern and
the heat exchange.

Figure\ref{fig:temp-distr} shows the spatial distribution of the Nusselt number
around the cylinders at $s/d=3.6$ and $s/d=4$. Even if it represents only a
snapshot of an unsteady behaviour, it remarks the asymmetric spatial
distribution of the local Nusselt number according to the oscillatory flow
field. At $s/d=3.6$, fig.\ref{fig:temp-distr}a, the Nusselt number distribution
shows that the cylinders $3$, $4$, $5$ and $6$ have an alternate position of
the maximum on the cylinder surface. Whereas at $s/d=4$,
fig.\ref{fig:temp-distr}b, the maxima of the local Nusselt number are located in the same position. 

Thus, the vortices affect the heat transfer mechanisms on the cylinders due to
their ability to carry cold fluid near the cylinders. According to the vortex
shedding the heat exchange conditions change, therefore the time series of the
mean Nusselt number experience an oscillating behaviour too, as showed in
fig.\ref{fig:nu-ts} (fig.\ref{fig:nu-ts}a at s/d = 3.6, fig.\ref{fig:nu-ts}b
at s/d = 4).  In figure \ref{fig:nu-ts} the Nusselt number of the cylinder
$1$ is almost constant for both the spacings. The cylinder $2$ has a higher
Nusselt number for $s/d=4.0$ because of the entrainment of cold fluid due to
its counter phase shedding respect to cylinder $1$, otherwise at $s/d=3.6$ the
vorticity pattern avoid the entrainment of cold fluid decreasing the its heat
exchange with the surrounding fluid.  For $s/d=4$ the following cylinders,
from $3$ to $6$, continue to decrease their Nusselt number monotonically
according to the in phase vortex shedding.  Decreasing the spacing ratio at
$3.6$, the Nusselt number of cylinder $3$ increases its value respect to
cylinder $2$.  Also the following cylinders keep their values significantly
higher than the case with $s/d=4$ according to the counter phase vortex
shedding that causes cold fluid entrainment in the gap regions. 

\begin{figure}[h!] 
\centering
\includegraphics[width=0.65\textwidth]{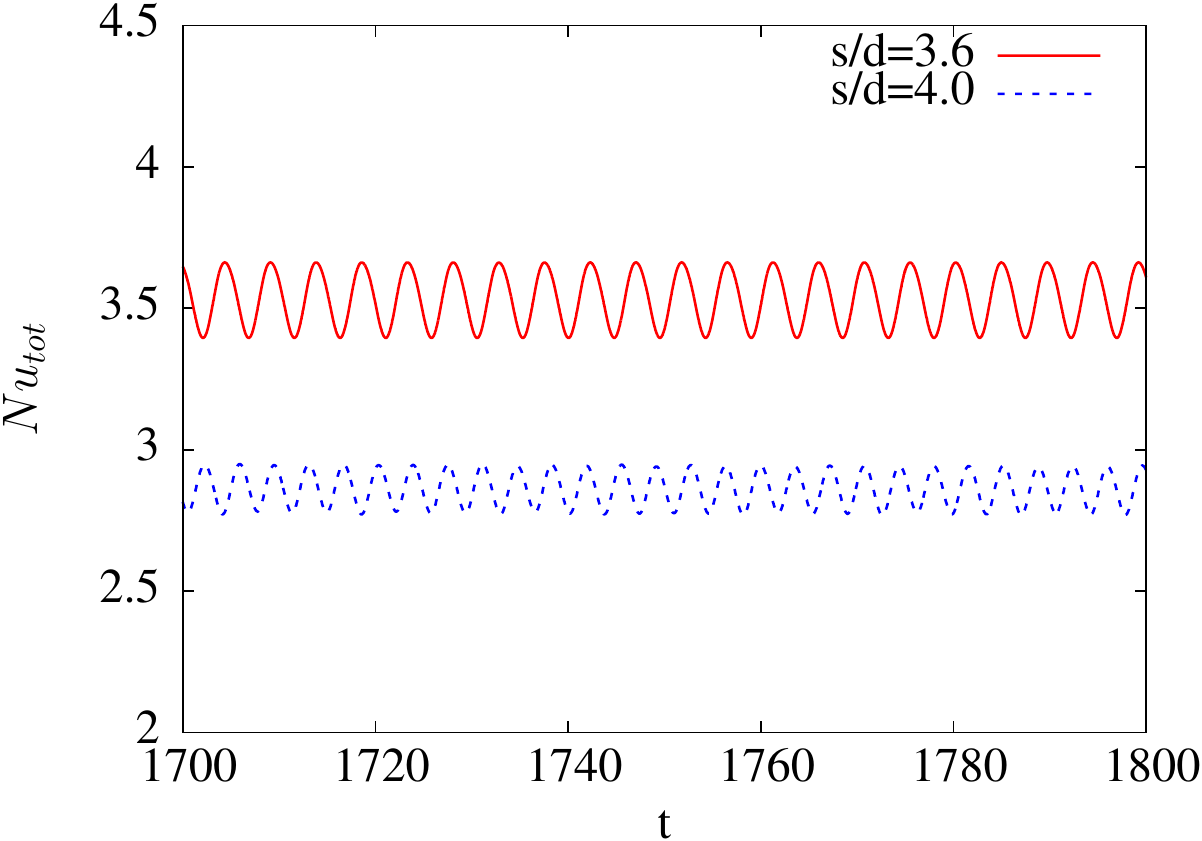} 
\caption{
Time series of the total mean Nusselt number, $Nu_{tot}=(\sum_{\imath=1}^{6}Nu_\imath)/6$, 
of the six in-line cylinders for a spacing ratio of $3.6$ and $4$.
(Colour online).}\label{fig:nu-total} 
\end{figure}

Although the instantaneous temperature and flow  distributions are useful in
the qualitative description of the phenomenon, in order to characterize this
problem, an analysis on the time-average quantities is needed.  Figure
\ref{fig:nu-ts}a shows the time average Nusselt numbers of each cylinder for
$s/d = 3.6$ and $s/d = 4$. Figure \ref{fig:nu-split}a confirms that there is
the transition in the heat transfer of the cylinder array between a spacing
ratio of $3.6$ and $4$.  For each time series of the Nusselt number, the main
frequency and the phase shift, respect to the first cylinder phase, have been
found. The results are reported in fig.\ref{fig:nu-split}b in a polar diagram
where the distance from the origin represents the mean Nusselt, whereas the
angular position is the phase shift respect to the first cylinder. In order to
measure the phase shifts, we verified that all the cylinders have the same
main frequency. According to the power spectra of the mean Nusselt number of
each cylinder, the heat transfer signal has one main frequency equal to $0.21$
and $0.279$ for a value of the spacing ratio of $3.6$ and $4$, respectively.
Looking at the time series of the mean Nusselt number for each cylinder,
fig.\ref{fig:nu-ts} and fig.\ref{fig:nu-split}b, qualitatively, the phase
shift between the different signals seems to vanish changing the spacing ratio
from $3.6$ to $4$. The phase shifts at $s/d=3.6$ are homogeneously distributed
in the polar diagram respect to $s/d=4$, for which the most part of the
cylinders has the same phase shift. It suggests that the phase shifts of the
mean Nusselt number can be related to the heat exchange efficiency. Indeed
this results find a theoretical approach about the mechanism of counter phase
shedding: the counter phase shedding causes a phase shift of the mean Nusselt
number, while an in phase shedding leads to a zero phase shift of the mean
Nusselt number. It is worth noting that the
$Nu_{tot}=(\sum_{\imath=1}^{6}Nu_\imath)/6$ for a spacing ratio of $3.6$ is
about $25\%$ higher than those at a spacing ratio of $4$
(fig.\ref{fig:nu-total}).  

% According to
%the flow field, the time series of the mean Nusselt number experiences an
%oscillating behaviour too as showed in fig.\ref{fig:nu-ts}. Indeed the change
%in the spacing ratio affects the heat transfer response of the array in terms
%of both the time-average cylinder Nusselt numbers (fig.\ref{fig:nu-split}a)
%and the relative phase shift between the Nusselt time-series
%(fig.\ref{fig:nu-split}b).  

%Indeed this
%result confirms that the phase shifts of the mean Nusselt number of each
%cylinder give a sign about the heat transfer efficiency of the array. 

\section{Conclusions} The numerical simulations of the flow around six in-line
iso-thermal circular cylinders at $Re=100$ for two values of the spacing ratio
($3.6$ and $4$) is presented. Between the spacing ratios $3.6$ and $4$ a flow
pattern transition is found. Long term and large domain simulations reveals at
both the spacing ratio ($3.6$ and $4$) there are a main frequency and one high
secondary frequency in the amplitude spectra of the lift coefficient. In
particular, for a spacing ratio of $4$, one low secondary frequency has been
found near the cylinder $6$ (see fig.\ref{fig:fft-six-4}). This lowest
secondary frequency depends on a secondary vortex street in the far wake
region. 

Two new flow modes are classified for six in-line circular cylinders
configuration at a $Re=100$.  Stable shear layer (SSL) mode and shear layer
secondary vortices (SLSV) mode.  The former occurs at $3.6$ spacing ratio,
where the shedding vortices coming from the downstream cylinder merge in two
stable parallel shear layer rows that are dissipated due to viscosity.  At a
spacing ratio of $4$ the flow is organized in the SLSV mode for which the
shear layers become unstable in the far wake region producing a secondary
vortex street.  The SLSV mode provides clear evidence that the drag is reduced
up to $30\%$ with respect SSL mode. 

The overall heat exchange is heavily affected by the spacing ratio at the
transition. In fact, at $3.6$ spacing ratio, the flow pattern induces the
mixing of cold flow near the heated cylinders $3$, $4$, $5$ and $6$ increasing
their local Nusselt numbers respect to the case at a spacing ratio of $4$.
For the same reason, cylinder $2$ has an higher Nusselt number for a spacing
ratio of $4$ respect to the $3.6$ case. The oscillating dynamics of the time
series of the local Nusselt number states that the higher phase shift between
the different cylinder, the higher heat exchange on the cylinder is. It is
worth noting that the overall heat exchange increase of $25\%$ reducing the
spacing ratio from $4$ to $3.6$.

\section{Ackonwledgement} We acknowledge the kind support provided by the IT
staff (M. Franza, A.  Chiffi, S. Sergiampietri, E. Rizzo) of the ``Centro
Cultura Innovativa d'Impresa'', University of Salento (Italy), where the
computations were carried out.

\bibliography{bibliography}

\end{document}